\documentclass[
onecolumn, superscriptaddress,prd,tightenlines,
nofootinbib,eqsecnum,
amsfonts,amsmath,amssymb]{revtex4}

\usepackage{bm}
\usepackage{graphics,graphicx,epsfig,color,ulem}
\usepackage{subfigure}

\begin{document}

\title{Reply to the comment on:\\
``Does an atom interferometer test the gravitational redshift\\ 
at the Compton frequency\,?''}

\author{Peter Wolf} \affiliation{LNE-SYRTE, CNRS UMR 8630, UPMC,
  Observatoire de Paris, 61 avenue de l'Observatoire, 75014 Paris,
  France} \author{Luc Blanchet} \affiliation{GRECO, Institut
  d'Astrophysique de Paris, CNRS UMR 7095, UPMC, 98$^\text{bis}$
  boulevard Arago, 75014 Paris, France} \author{Christian J. Bord\'e}
\affiliation{LNE-SYRTE, CNRS UMR 8630, UPMC, Observatoire de Paris, 61
  avenue de l'Observatoire, 75014 Paris, France}
\affiliation{Laboratoire de Physique des Lasers, CNRS UMR 7538, 99
  avenue Jean-Baptiste Cl\'ement, 93430 Villetaneuse, France}
\author{Serge Reynaud} \affiliation{Laboratoire Kastler Brossel, CNRS
  UMR 8552, ENS, UPMC, Campus Jussieu, 75252 Paris, France}
\author{Christophe Salomon} \affiliation{Laboratoire Kastler Brossel
  et Coll\`ege de France, CNRS UMR 8552, ENS, UPMC, 24 rue Lhomond,
  75231 Paris, France} \author{Claude Cohen-Tannoudji}
\affiliation{Laboratoire Kastler Brossel et Coll\`ege de France, CNRS
  UMR 8552, ENS, UPMC, 24 rue Lhomond, 75231 Paris, France}
\date{\today}

\begin{abstract}
Hohensee, Chu, Peters and M\"uller have submitted a comment (arXiv:1112.6039 [gr-qc]) on our paper ``Does an atom interferometer test the gravitational redshift at the Compton frequency\,?'', Classical and Quantum Gravity {\bf 28}, 145017 (2011), arXiv:1009.2485 [gr-qc]. Here we reply to this comment and show that the main result of our paper, namely that atom interferometric gravimeters do not test the gravitational redshift at the Compton frequency, remains valid.
\end{abstract}

\maketitle

In their comment Hohensee, Chu, Peters and M\"{u}ller (referred to
as HCPM here) \cite{HCPM} re-affirm their earlier interpretation, based on an
analogy with classical clock experiments, of atom interferometric
gravimeters as testing the gravitational redshift. They repeat
previous arguments, in particular the cancellation between the
time dilation and laser phase contributions to the phase shift,
and arguments centered around Schiff's conjecture. Here we
re-affirm that these arguments are incorrect or irrelevant, as
shown below, and that the main result of our paper \cite{WolfCQG},
namely that atom interferometric gravimeters do not test the
gravitational redshift at Compton's frequency, remains valid.

First HCPM argue that the absence of the Compton frequency (or
equivalently the atom's mass) in the final phase shift, does not
prevent the experiment from being a redshift test at that
frequency. They claim that this is also the case for redshift
tests using standard clocks, like the Pound-Rebka experiment,
where the redshift effect is cancelled by a controlled first order
Doppler effect, thus effectively leaving a null signal independent
of the 14.4 keV transition frequency. This analogy is fallacious~:
the Pound-Rebka experiment is based on a resonance between atoms
sitting at the top and bottom of a tower. The controlled Doppler
shift is just a clever experimental method for scanning the
resonance which is affected by the redshift (cf. Figure 1. in
\cite{PoundSnider}). The resulting signal is thus null on the
exact center of the resonance curve but not at other frequencies.
In the atom gravimeter in contrast, the cancellation of the
Compton frequency is inherent to the interferometer~: the action
phase shift involving the Compton frequency is always zero as
shown in \cite{WolfCQG} and the measurement of that frequency (and
of the associated redshift) is intrinsically impossible.
Furthermore it is manifestly incorrect to state as HCPM do that
the cancellation of the clock frequency is a generic feature of
any redshift measurement. Most clock experiments
\cite{Vessot,Cacciapuoti} measure frequencies of two
electromagnetic signals delivered by two clocks. The redshift
effect is thus associated to a relation such as $\Delta f = f_0
\Delta U/c^2$ where $\Delta f$ is the measured frequency
difference and $f_0$ the common proper frequency of the clocks.
This could also be done in principle for a Pound-Rebka-like
experiment if experimental means (say frequency combs) were
available for accurate measurements of frequencies in the X-ray
domain. By contrast, in the atomic interferometer considered by
HCPM, there is no real emission of a photon at the Compton
frequency and no measurement of the frequency of this photon by a
detector.

The interferometer considered by HCPM is in fact not an atomic
clock. The two states $g$ and $g'$ appear symmetrically in the two
arms of the interferometer and no resonance is obtained when one
varies the laser frequency. As explained in \cite{WolfCQG}, there
are other non-symmetric atomic interferometers where the output
signal exhibits resonances when the laser frequency is varied.
They provide a measurement of the atomic frequency $\omega_{gg'}$
(see discussion after (2.21) in \cite{WolfCQG}), so that these
interferometers can be also considered as atomic clocks
oscillating at $\omega_{gg'}$. To get an oscillation at the
Compton frequency, one would need to have in the two arms of the
interferometer two states, whose energies differ by $mc^2$, which
is far beyond present day technology. Furthermore, to test the red
shift at Compton frequency, two different clocks of this type
located at two different altitudes would have to be built and
their frequencies compared.

In a related argument (eq. (1) and text below in their comment),
HCPM restate their previous claim that in the total phase shift
the time dilation phase $\phi_t$ cancels the laser phase
$\phi_{\rm laser}$ and not the gravitational phase $\phi_r$,
whereas we have demonstrated in \cite{WolfCQG} that the free
evolution phase $\phi_S=\phi_r+\phi_t$ is necessarily zero for all
closed paths that follow trajectories obtained from the same
Lagrangian as the one used for the phase integral. The latter fact
implies that the difference of action integrals is zero and that
the Compton frequency and redshift at that frequency are
unmeasurable, no matter what modified Lagrangian is used (i.e.
independently of any redshift violating parameter $\beta$). We
re-affirm that the fundamental cancellation is between $\phi_r$
and $\phi_t$, one reason being that $\phi_S=\phi_r+\phi_t=0$ is an
invariant statement, since $\phi_S$ is the contour integral of the
proper time (see first term of the right-hand side of (1.3) in
\cite{WolfCQG}) which is invariant under coordinate
transformations. By contrast the statement that $\phi_t+\phi_{\rm
laser}=0$ is not invariant and is only true in the rest frame of
the lasers. To see this in a simple example, analyze the atom
gravimeter in a frame that is freely falling in the Earth's
gravitational field. An elementary calculation shows that $\phi_r$
and $\phi_t$ are separately zero in that frame, $\phi_r=\phi_t=0$,
but that $\phi_{\rm laser}= k\,g\,T^2$, where $k$ is the laser
wavevector and $T$ the interrogation time, because the lasers are
accelerated upwards in that frame. Thus there is no cancellation
between $\phi_t$ and $\phi_{\rm laser}$ in that frame but still we
have $\phi_S=\phi_r+\phi_t=0$.

Concerning Schiff's conjecture, we do not agree with the analysis
of HCPM. Schiff's conjecture states that any violation of the
universality of free fall (UFF) also implies a violation of the
universality of clock rates (UCR) and vice versa. Does this mean
that it makes no sense to distinguish between the two types of
tests~? Our answer is ``No'', because the relation between UFF and
UCR tests depends on the experiments under consideration, on the
nuclear or atomic models used for the analysis of the clocks and
the test masses, and on the particular model used for the
violation of UFF and UCR.

The next question is then whether the atom gravimeter measurement
is more directly comparable to UFF or UCR tests~? As shown in
\cite{WolfCQG} but also e.g. in \cite{Giulini} the answer is
clearly UFF as it constrains the same parameters as other UFF
tests, whereas it requires nuclear and atomic models to relate it
to parameters measured in UCR tests. To make this point more
explicit, let us quote the SME analysis presented in
\cite{Hohensee}~: the UFF test of \cite{Schlamminger} sets a limit
on $\beta^{\rm Be}-\beta^{\rm Al}$ (to leading order), the atom
gravimeter test sets a limit on $\beta^{\rm Cs}-\beta^{\rm grav}$
(where ``grav'' stands for the test mass in the classical
gravimeter) and the Pound-Rebka experiment on $\xi^{\rm
Mossb}_{\rm ^{57}Fe}-\beta^{\rm grav}$. To relate the former two
one only requires the atomic composition of Be, Al and grav (equ.
(8) of \cite{Hohensee}) but to relate them to the Pound-Rebka
measurement one additionally requires a nuclear model for the 14.4
keV transition in $^{57}$Fe (see text before (11) in
\cite{Hohensee}).

To conclude, let us briefly recall the compelling argument (see
e.g. \cite{Giulini}) that atom interferometers test the UFF. This
argument is agreed on since the realization of the first atomic
gravimeters two decades ago \cite{Kasevich}. The atoms with
inertial mass $m_i$ and gravitational mass $m_g$ obey the
Lagrangian $L=\frac{1}{2}m_i\,\dot{z}^2 - m_g\,g\,z$ in the
gravitational field of the Earth. The phase shift is the sum of
the free evolution phase given by the difference of action
integrals $\phi_S$ along the two paths, and of the contribution
$\phi_{\rm laser}$ coming from the interaction with the lasers.
The free evolution phase $\phi_S$ is exactly zero for a closed
total path (see e.g. \cite{WolfCQG}), and the phase shift reduces
to $\phi_{\rm laser}=(m_g/m_i)\,g\,k\,T^2$ using the position of
the atoms deduced from the Lagrangian. Comparing with the
measurement by a nearby classical gravimeter of the gravitational
acceleration of a freely falling macroscopic mass (corner cube),
$g_{\rm meas}=(M_g/M_i)g$, one obtains to first order $\phi_{\rm
laser}=(1+\eta)\,g_{\rm meas}\,k\,T^2$ where
$\eta=m_g/m_i-M_g/M_i$ is the E\"otvos parameter between the atoms
and the corner cube, showing that measurement of the phase shift
gives a test on possible violations of the UFF between atoms and
macroscopic masses.

\end{document}